\documentclass[RNAAS]{aastex63}

\newcommand{\frborig}{FRB\,20200120E}

\newcommand{\UCB}{Department of Astronomy,  University of California Berkeley, Berkeley, CA}

\newcommand{\SETI}{SETI Institute, Mountain View, California}
\newcommand{\KZA}{University of Malta, Institute of Space Sciences and Astronomy}

\newcommand{\mcgill}{Department of Physics, McGill University, 3600 rue University, Montr\'eal, QC H3A 2T8, Canada}
\newcommand{\msi}{McGill Space Institute, McGill University, 3550 rue University, Montr\'eal, QC H3A 2A7, Canada}
\newcommand{\caltech}{Division of Physics, Mathematics, and Astronomy, California Institute of Technology, Pasadena, CA 91125, USA}
\newcommand{\msif}{McGill Space Institute Fellow}
\newcommand{\frqnt}{FRQNT Postdoctoral Fellow}
\newcommand{\DI}{Dunlap Institute for Astronomy \& Astrophysics, University of Toronto, 50 St.~George Street, Toronto, ON M5S 3H4, Canada}
\newcommand{\UoM}{Jodrell Bank Centre for Astrophysics, Department of Physics \& Astronomy, Alan Turing Building, The University of Manchester, M13 9PL, United Kingdom}

\begin{document}

\title{Absence of bursts between 4 -- 8 GHz from \frborig\ located in an M\,81 Globular Cluster}

\correspondingauthor{Vishal Gajjar}
\email{vishalg@berkeley.edu}

\author[0000-0002-8604-106X]{Vishal Gajjar}
\affiliation{\UCB}

\author[0000-0002-2551-7554]{Daniele Michilli}
\affiliation{\mcgill}
\affiliation{\msi}

\author[0000-0001-9855-5781]{Jakob T. Faber}
\affiliation{Department of Physics and Astronomy, Oberlin College, Oberlin, OH}
\affiliation{\UCB}

\author[0000-0002-4064-7883]{Sabrina Berger}
\affiliation{\mcgill}
\affiliation{\msi}

\author[0000-0003-4823-129X]{Steve Croft}
\affiliation{\UCB}
\affiliation{\SETI}

\author[0000-0002-8912-0732]{Aaron B. Pearlman}
\affiliation{\mcgill}
\affiliation{\msi}
\affiliation{\caltech}
\affiliation{\msif}
\affiliation{\frqnt}

\author[0000-0003-3154-3676]{Ketan R. Sand}
\affiliation{\mcgill}
\affiliation{\msi}

\author[0000-0002-7374-7119]{Paul Scholz}
\affiliation{\DI}

\author[0000-0003-2828-7720]{Andrew P. V. Siemion}
\affiliation{\UCB}
\affiliation{\SETI}
\affiliation{\UoM}
\affiliation{\KZA}


\begin{abstract}
We report the non-detection of dispersed bursts between $4 - 8$\,GHz from 2.5 hours of observations of \frborig\ at 6\,GHz using the Robert C.~Byrd Green Bank Telescope. Our fluence limits are several times lower than the average burst fluences reported at 600 and 1400\,MHz. We conclude that these non-detections are either due to high-frequency bursts being weaker and/or scintillation-induced modulated. It is also likely that our observations were non-concurrent with any activity window of \frborig. 
\end{abstract}

\section{Introduction}
Fast radio bursts (FRBs) are millisecond-duration pulses of extragalactic origin. FRB\,121102A has been detected across a wide frequency range, from 400\,MHz \citep{chime2019R1} to 8\,GHz \citep{Gaj18}, giving insights into the burst emission properties \citep{mic18}; no other FRB has been detected across such a large range of frequencies. 
The newly-discovered \frborig\ was found to have a dispersion measure (DM) of 87.8\,pc\,cm$^{-3}$, which is the lowest reported among all FRBs \citep{M81FRB_BGK21}. Recently, \cite{M81_GC_loc} announced a precise localization of three bursts from \frborig\ coinciding with a globular cluster in the M81 system. Thus, \frborig\ provides an excellent opportunity to conduct deep observations at higher frequencies ($> 4$\,GHz). 
A detection at higher frequencies would offer the opportunity to (a) compare frequency-dependent burst rates, (b) measure scatter-independent rotation measure and polarization position angle, and (c) investigate frequency-dependent burst widths, sub-burst components, and downward-drifting behavior. Here, we report observations of \frborig\ from the Green Bank Telescope (GBT), using the C-band receiver at $4 - 8$\,GHz.

\section{Observations and analysis} 
We have an ongoing observing campaign with the GBT to trigger C-band observations for any repeating FRB reported by the CHIME/FRB project\footnote{\url{https://www.chime-frb.ca/repeaters}} with localization uncertainty smaller than the GBT beam ($2\farcm5$). The observations of \frborig\ were taken for $2.5$\,h. We utilized the Breakthrough Listen (BL) backend, which is a state-of-the-art 64-node GPU cluster \citep{MacMahon2018}, deployed primarily to conduct the most comprehensive search for evidence of intelligent life in the Universe \citep{Worden2017}. 
Recorded baseband voltages were converted to SIGPROC-formatted filterbank files \citep{Lebofsky2019}. As FRBs are known to show limited spectral coverage \citep{Gaj18,zhang18}, we divided our 4\,GHz  band into eight 500\,MHz-wide sub-bands, searched independently to improve the chance of detecting spectrally-limited bursts \citep{Faber_121102}. We used SPANDAK, a similar search pipeline to that described by \citet{Gaj18}, to blindly search for single pulses, with search parameters listed in Table \ref{tab:obs_and_results}. 

\begin{table}[]
    \centering
    \begin{tabular}{cccc}
    \hline
    \multicolumn{4}{c}{Observations and Search Parameters} \\
    \hline
    \hline
        Source Name & \frborig & \hspace{1cm} DM$_{\rm range}$   & $0 - 1000$\,pc\,cm$^{-3}$  \\
        RA(J2000) & 09$^{h}$57$^{m}$56.7$^{s}$  & \hspace{1cm} Widths  & $0.3 - 76$\,ms \\
        Dec(J2000) & $+68\degr 49\arcmin 31\farcs8$  & \hspace{1cm} S/N            &  $> 10$ \\
        Frequency &  $4 - 8$\,GHz      & \hspace{1cm} F$_{\rm UL}^{\rm full band}$  & 35\,mJy\,ms   \\
        T$_{\rm sampling}$ & $\sim 350\,\mu$s     & \hspace{1cm} F$_{\rm UL}^{\rm sub-band}$ ($\Delta\nu\sim 500$\,MHz)  &     100\,mJy\,ms  \\
        Obs. MJD (59259+) &  $0.969687500 - 1.072951388$  & \hspace{1cm} burst-rate$_{\rm UL}$  & $< 0.4$\,hr$^{-1}$ \\
        \hline
    \end{tabular}
    \caption{Summary of observations and single pulse searches of \frborig\ across $4 - 8$\,GHz. The F$_{\rm UL}^{\rm full band}$ and F$_{\rm UL}^{\rm sub-band}$ stand for minimum detectable fluence assuming a 1\,ms-wide top-hat burst for the full band and any one of the sub-bands, respectively.}
    \label{tab:obs_and_results}
\end{table}

\section{Results and conclusion}
We did not detect any significant dispersed burst across the full band ($4 - 8$\,GHz) nor in any of the eight sub-bands. Upper limits on the fluences are listed in Table \ref{tab:obs_and_results}. \cite{M81FRB_BGK21} reported an average burst fluence $\sim 2210$\,mJy\,ms at 600\,MHz. \cite{M81_GC_loc} reported multiple bursts at 1.4\,GHz with an average burst fluence $\sim 500$\,mJy\,ms and burst rate $\sim 0.68$\,hr$^{-1}$.  These fluences and burst rates are higher than our current fluence limits and our inferred burst rate. However, it is plausible that bursts at 6\,GHz are intrinsically weaker (or absent) than those detected at lower frequencies. \cite{Nimmo_2021_M81_microstruct} and \cite{majid_2021_M81_microstruct} reported that bursts from \frborig\ exhibit sub-burst structure at $< 100$\,ns scales, similar to `nanoshots' seen in giant pulses (GPs) from the Crab pulsar. 
\cite{Hankins_Crab_GP} compared simultaneous detections of several hundred GPs from the Crab pulsar at 1.4 and 5\,GHz and found the average instantaneous spectral index to be steep ($\alpha\sim -2$). 
If the bursts from \frborig\ are indeed similar to GPs from the Crab pulsar, such a steep spectrum would render burst fluences at 6\,GHz below our detection threshold. Furthermore, the expected diffractive scintillation bandwidth varies from 200\,MHz to 3.5\,GHz across our observed band towards the source (see \citealt{majid_2021_M81_microstruct}). If the bursts are spectrally narrower than this scintillation bandwidth then they are likely to be 100\% modulated, which could yield non-detections. It is also possible that \frborig\ exhibits enhanced burst activity windows (seen from other repeating FRBs; \citealt{zhang18,price18_R1_nondetection}) and our observations were non-concurrent with such active phases. However, in the absence of any other reported non-detection, we are unable to verify if \frborig\ indeed exhibits such prolonged inactive phases. Future simultaneous multi-frequency multi-epoch observations may provide insights into the burst properties around 6\,GHz. 

\acknowledgments
Breakthrough Listen is managed by the Breakthrough Initiatives, sponsored by the Breakthrough Prize Foundation. The Green Bank Observatory is a facility of the National Science Foundation, operated under cooperative agreement by Associated Universities, Inc. J.T.F. and S.C. were supported by the National Science Foundation under Grant No.~1950897. D.M. is a Banting Fellow. A.B.P is a McGill Space Institute (MSI) Fellow and a Fonds de Recherche du Quebec - Nature et Technologies (FRQNT) postdoctoral fellow. P.S. is a Dunlap Fellow and an NSERC Postdoctoral Fellow. The Dunlap Institute is funded through an endowment established by the David Dunlap family and the University of Toronto.

\bibliographystyle{aasjournal}
\bibliography{mybib_frb,ao-Cband,frbmodels,rnaas}

\end{document}